# Data Compaction –
# Compression without Decompression

Steffen Görzig

*Daimler AG*
*HPC G012*
*71059 Sindelfingen, Germany*
*steffen.goerzig@daimler.com*

*Abstract:* Data compaction is a new approach for lossless and lossy compression of read-only array data. The biggest advantage over existing approaches is the possibility to access compressed data without any decompression. This makes data compaction most suitable for systems that could currently not apply compression techniques due to real-time or memory constraints. This is true for the majority of all computers, i.e. a wide range of embedded systems.

## 1. Introduction

Data compression is a well-known approach for many years and is successfully applied in various use cases, e.g. to reduce data storage space or to increase data transmission capacity. However, all current compression techniques share the same drawback: compressed data must be decompressed to be used and decompression imposes extra computational or memory costs. On a typical personal computer these costs are acceptable. But a wide range of embedded systems cannot use data compression due to
- Real-time constraints, e.g. when certain startup times have to be ensured.
- Memory limitations. For example when only a few kilobytes are available the additional memory consumption for the decompressed data as well as for the decompression algorithm itself is too large. Additional memory chips would increase the production costs significantly, since embedded systems are typically sold in large quantities.

This paper introduces data compaction which addresses this problem. Similar techniques are known for file systems [1] or for code compaction [2].

## 2. Data Compaction

Data compaction is a new approach for lossless and lossy compression of read-only array data. Data compaction is:
- An efficient way to access compressed data without any decompression.
- No source code modification is needed to access compacted data.
- A technology that can be integrated transparent into the compiling process.

## 2.1. In a Nutshell

Data compaction basically consists of three steps. The input data are read-only arrays of any sizes, dimensions, and types. In the first step all input arrays are converted to arrays of a basic type, for current computers byte is a good choice. The second step is to build a minimal sized one-dimensional array of the basic type byte containing all transformed input data byte arrays. The third step is to replace the input arrays by pointers into the one-dimensional array with the corresponding input array data.

Here is a simple example assuming a platform using two's complement for negative values with two bytes big-endian representation for integer. The input array `iA` is a two-dimensional array of type integer, `ucA` is a one-dimensional array of type byte:

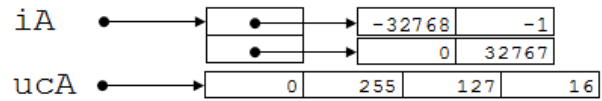

In the first step the data is converted to byte arrays using the above platform assumptions (e.g. "-1" is converted to the bytes "255 255"):

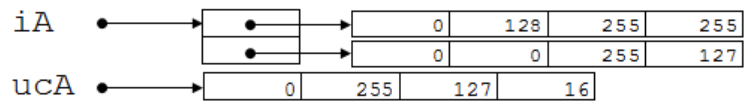

The second step is the compaction of the data: to build the one-dimensional byte array of minimal size containing all input data:

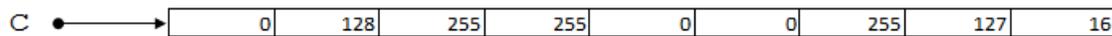

The last step is to replace the input arrays by pointers to the compacted data:

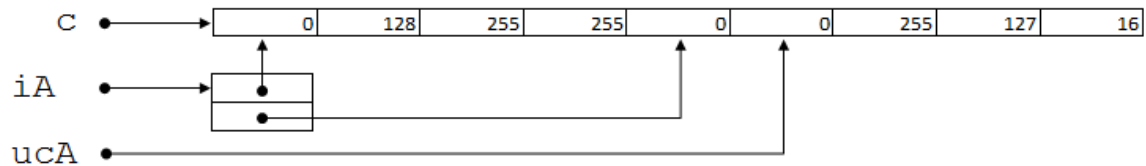

In this example three bytes are saved by data compaction in comparison to the original data. The realization of data compaction depends on the used programming language. In the following the given example is implemented in the programming language C, since this language is widely used for embedded systems:

```
const int iA[2][2] = {{-32768,-1}, {0, 32767}};
const unsigned char ucA[4] = {0, 255, 127, 16};
…
/* data access */
i = iA[0][1];
uc = ucA[1];
```

### Step 1: Input Transformation:
```
const unsigned char iABytes[4][4] = {{0, 128, 255, 255}, {0, 0, 255, 127}};
const unsigned char ucABytes[4] = {0, 255, 127, 16};
```

### Step 2: Compaction:
```
const unsigned char c[9] = {0, 128, 255, 255, 0, 0, 255, 127, 16};
```

### Step 3: Output Generation:
```
const int *iA[2] = {(const int*)&c[0],(const int*)&c[4]};
const unsigned char *ucA = (const unsigned char*)&c[5];
…
/* data access */
i = iA[0][1];
uc = ucA[1];
```

The C example shows that the source code for accessing the data has not to be modified for compaction. Moreover no programmer interaction is needed assuming compaction is integrated in the compilation process – steps 1-3 could be done automatically by an additional compilation step.

## 2.2. Step 1: Input Transformation

The purpose of input transformation is to build a set of one-dimensional byte arrays. This set is handed over to step 2, compaction. One-dimensional arrays can be processed directly; multi-dimensional arrays must be decomposed to their one-dimensional parts. The transformation of the one-dimensional input data to byte arrays depends on the internal platform representation of data types. In C for example the following aspects have to be considered:
- Number of bytes for a specific type (e.g. two or four bytes for integer).
- Endianness (e.g. little-endian or big-endian).
- Binary representation of negative values (e.g. two's complement).

With the internal representation defined, the transformation for each data type is straight forward (see example above). If this step should be performed automatically (e.g. by a preprocessor) there are several ways to identify the read-only input data arrays:
- Data-flow analysis is used to determine read-only arrays.
- The programmer can tell the preprocessor which array variables to process.
- The computer language has a keyword for read-only data arrays (e.g. `const` in C).

## 2.3. Step 2: Compaction

The input for the compaction step is the set of one-dimensional byte arrays provided by the transformation step. The core of data compaction is to reduce this array set. This could be again a set of one-dimensional byte arrays, for practical reasons this set is reduced to only one array. So the task is to build a single one-dimensional byte array of minimal or close to minimal size containing all input data. A part of this optimization task is addressed by the shortest superstring problem:

Given a set of strings $S=\{s_1, s_2, \ldots, s_n\}$ over a finite alphabet $\sum$, a *superstring* of $S$ is a string that contains each $s_i$ as a contiguous substring. The *shortest superstring* problem is to find a superstring of minimum length. Since this problem is NP-complete approximation algorithms are applied. The best known proven approximation factor is $2\frac{1}{2}$ [3, 4]. However, a straightforward greedy approach is conjectured to give an approximation ration of 2.

In 2.3.1. and 2.3.2. already known approximation methods are shown to give a better understanding how this works. New methods for compaction are shown in 2.3.3.-2.3.5. The methods 2.3.3. and 2.3.5. may also create additional computational costs and should be avoided if the goal is no cost at all. All methods generate lossless compression except for the lossy compression method 2.3.4. (and 2.3.4. in combination with 2.3.5.). When a method produces a set of arrays, the arrays are concatenated to a single one-dimensional byte array afterwards.

### 2.3.1. Remove Sub-Arrays

If any array is completely part of another array, it is removed. Example:
Byte arrays
```
{0, 16, 32}
{0, 16, 32, 128}
{1, 17}
{17}
```
Resulting arrays after removing sub-arrays:
```
{0, 16, 32, 128}
{1, 17}
```

### 2.3.2. Greedy Algorithm

Determine the two most overlapping arrays and replace them with the array obtained by overlapping them. Repeat this until only one array is left. Example:
Byte arrays:
```
{0, 16, 155}
{155, 17, 0, 16}
{155, 233, 0}
```
Calculate number of overlapping data for each combination:
```
Combination {0, 16, 155} {155, 17, 0, 16} : overlap {155}
Combination {155, 17, 0, 16} {0, 16, 155} : overlap {0, 16}
Combination {0, 16, 155} {155, 233, 0}    : overlap {155}
Combination {155, 233, 0} {0, 16, 155}    : overlap {0}
Combination {155, 17, 0, 16} {155, 233, 0}: overlap {}
Combination {155, 233, 0} {155, 17, 0, 16}: overlap {}
```
Concatenate the best overlap result and remove the overlap input arrays:
```
Concatenate {155, 17, 0, 16} {0, 16, 155} to {155, 17, 0, 16, 155}
Remove {155, 17, 0, 16} and {0, 16, 155}
```
Resulting byte arrays after one step:
```
{155, 17, 0, 16, 155}
{155, 233, 0}
```
Resulting byte array after two steps:
```
{155, 17, 0, 16, 155, 233, 0}
```

The greedy algorithm can be stopped when there is no new overlap anymore.

### 2.3.3. Mapping functions

If an array can be mapped to another array (or part of that array) by a function, it can be replaced by only keeping the mapping function. Hence in general mapping functions are hard to find automatically, this might be only useful for special cases (i.e. for lookup tables with different scale). The mapping function has to be taken into account when accessing the data, this causes additional computational costs. Example:
Mapping Function $f(x) = x/2$:
Byte arrays:
```
{0, 16, 32}
{0, 8}
```
Resulting array:
```
{0, 16, 32}
```

### 2.3.4. Merge Similar Arrays

For lossy compression similar arrays can be merged. This can be done by calculating the distance between byte arrays (respectively between each sub-array/array combination for arrays of different length). If the result is below a predefined threshold value the arrays can be merged. The merge method should be adapted to the type of the array data.

As example a simple distance function is given by: $\frac{\sum_{i=1}^{n}(|x_i - y_i|)}{n}$

The threshold value is set to 10. Byte arrays:
```
{0, 16, 32}
{0, 14}
```
Calculated distances:
```
{0, 14} and {0, 16} : distance 1
{0, 14} and {16, 32}: distance 17
```
Merge arrays below threshold using arithmetic mean for each value:
```
{0, 14} and {0, 16} to {0, 15}
```
Resulting array:
```
{0, 15, 32}
```

Of course lossy compression should only be applied for use cases that are able to handle these kinds of data.

### 2.3.5. Reverse Arrays

Apply 2.3.1.-2.3.4. with reverse arrays in addition. The access to the data must be inverted in case the input array was inverted during compaction. If the preprocessor replaces access indices (e.g. plus by minus for index calculations) this causes no computational costs, otherwise it does and must be handled during runtime. The consideration of reversed arrays does not increase the asymptotic worst-case complexity for the compaction process. Example:

Byte arrays:
```
{0, 16, 32}
{32, 16, 0}
{0, 17}
```
Resulting array applying 2.3.1. and 2.3.2. with reverse arrays:
```
{32, 16, 0, 17}
```

## 2.4. Step 3: Output Generation

The output generation replaces the input arrays by pointers to the one-dimensional byte array provided by the compaction step. This step strongly depends on the used programming language. However typically the following points have to be taken into account:
- The conversion to the types of the input data.
- Additional array pointer variables for arrays with dimension > 2.
- Null pointers are added at the highest possible dimension allowing efficient handling of sparse matrices.

Replacing array variables by pointers or adding pointer variables may lead to additional memory consumption by the size of the pointer representation. Therefore programs can get larger if compaction has no or little effect.

## 2.5. Restrictions

Data compaction is a transformation which can cause restrictions when used with real programming languages. In the following this is explained for the language "C". If data compaction would be part of a compiler, most of the mentioned points could be handled by the compiler itself. Some points must be handled by the programmer or depend on the used hardware.

As show with type `int`, the input transformation must be aware of compiler or platform dependent types (e.g. for `enum`, `wchar_t`, `struct`), padding bytes and for memory alignment.

In the "C" examples arrays are replaced by pointers. This is an indirection which can result to different machine code. Array element addresses cannot be calculated by start address plus offset anymore. Now there is a read access to the pointer variable. Since this is slower on typical hardware, this can be considered as decompression cost.

Pointers make no difference for the array subscript operator "[]". But they make a difference e.g. for the address operator "&". This is because the declaration between arrays and pointers differs, which has to be taken into account by programmers (e.g. functions expecting arrays as argument cannot be used anymore with pointers).

Since the input arrays are replaced by pointers to the compacted data, no data access must be allowed which assumes a contiguous representation of multi-dimensional arrays in memory when using compacted data:

```
/* uncompacted original: */
unsigned char arr[2][2] = {{1, 2}, {2 ,4}};
/* next data access only allowed for uncompacted code: */
unsigned char *arrPtr = &arr[0][0];
for (i=0; i<2*2; i++){
  printf("%u ", (*arrPtr));
  ++arrPtr;
}
/* compacted version: */
unsigned char c[3] = {1, 2, 4};
unsigned char *arr[2] = {(unsigned char*)&c[0], (unsigned char*)&c[1]};
int i, j;
/* for compacted and uncompacted code allowed: */
for (i=0; i<2; i++){
  for (j=0; j<2; j++){
    printf("%u ", uca[i][j]);
  }
}
```

Due to the same reason functions like `memcpy` should not be used to copy data from multi-dimensional compacted data to uncompacted data. Also functions like `sizeof` or `offsetof` should be handled with care. If data compaction should be performed automatically, it has to be ensured that this kind of data accesses does not occur. This can be achieved by data-flow analysis. Otherwise applying data compaction could result in incorrect programs. Analysis itself is hindered by the need of pointer analysis instead of simpler array analysis.

If the hardware uses bank switching for memory, the compacted data might not be accessible anymore in every context. Like with all compression algorithms, memory consumption is not projectable, since small changes can lead to very different compaction results. This also prevents a simple validation on binaries for small changes.

## 3. Implementation

As proof of concept the tool dataCompactor was implemented. The implementation is described in the next section. An example calculated by dataCompactor is given in 3.2., in 3.3. possible optimizations are mentioned.

## 3.1. dataCompactor

The tool dataCompactor was implemented as a python library. The input data must be provided as a python program using this library, the output is generated C-Code. Supported input data is:
- Read-only arrays of different sizes. Array dimensions might be between 1 and 3.
- C-types: `unsigned char`, `signed char`, `unsigned int`, and `int`.

The target platform can be specified by:
- Number of bytes for integer (typically two or four bytes).
- Little- and big-endian.
- Two's complement for binary representation of negative values.

The compaction step implements the methods "Remove Sub-Arrays" (2.3.1.) and "Greedy Algorithm" (2.3.2). For the greedy algorithm three strategies are implemented if there is more than one of maximum size overlap:
- Continue with the first found maximal overlap.
- Continue with the last found maximal overlap.
- Pick a random maximal overlap (seed for random numbers can be set).

The output generation can:
- Generate uncompacted C-Code as reference.
- Generate compacted C-Code including separate pointer variables for arrays with dimension >2 and NULL pointer handling.
- The variable name for the compacted data can be specified.
- Optional C keywords `static` and `const` can be generated.

## 3.2. Output Example

To provide a better understanding of how data compaction works in practice a larger example is given. Especially the pointers to the compacted data are shown in more detail making this part self-explanatory for programmers. This example is generated by dataCompactor using the methods "Remove Sub-Arrays" (2.3.1.) and "Greedy Algorithm" (2.3.2). Two's complement for negative values is used with two bytes big-endian representation for integer. The example includes the C-types `unsigned char`, `signed char`, `unsigned int`, and `int` with dimensions 1-3.

The uncompacted code:
```
const unsigned char arrayLen = 2;
const unsigned char uchar1DArr[2] = {2,4};
const signed char schar1DArr[2] = {-128,-64};
const unsigned char uchar2DArr[2][2] = {{2,4},{8,16}};
const signed char schar2DArr[2][2] = {{-128,-64},{-32,-16}};
const unsigned char uchar3DArr[2][2][2] = {{{2,4},{8,16}},{{32,64},{128,255}}};
const signed char schar3DArr[2][2][2] = {{{-128,-64},{-32,-16}},{{0,32},{64,127}}};
const unsigned int uint1DArr[2] = {4,16};
const int int1DArr[2] = {-32768,-4096};
unsigned int uint2DArr[2][2] = {{4,16},{64,256}};
const int int2DArr[2][2] = {{-32768,-4096},{-512,-64}};
const unsigned int uint3DArr[2][2][2] = {{{4,16},{64,256}},{{1024,4096},{16384,65535}}};
const int int3DArr[2][2][2] = {{{-32768,-4096},{-512,-64}},{{0,512},{4096,32767}}};
```

The same example with compacted code:
```
const unsigned char c[41] =
{0,4,0,16,0,254,192,255,64,127,0,0,2,4,0,32,64,0,0,1,0,128,0,240,0,16,255,127,0,64,255,255,128,192,2
24,240,8,16,128,255};
const unsigned char arrayLen = 2;
```

```c
const unsigned char *uchar1DArr = (const unsigned char*)&c[13];
const signed char *schar1DArr = (const signed char*)&c[33];
const unsigned char *uchar2DArr[2] = {(const unsigned char*)&c[13],(const unsigned char*)&c[37]};
const signed char *schar2DArr[2] = {(const signed char*)&c[33],(const signed char*)&c[35]};
const unsigned char *uchar3DArr0[2] = {(const unsigned char*)&c[13],(const unsigned char*)&c[37]};
const unsigned char *uchar3DArr1[2] = {(const unsigned char*)&c[16],(const unsigned char*)&c[39]};
const unsigned char **uchar3DArr[2] = {uchar3DArr0,uchar3DArr1};
const signed char *schar3DArr0[2] = {(const signed char*)&c[33],(const signed char*)&c[35]};
const signed char *schar3DArr1[2] = {(const signed char*)&c[15],(const signed char*)&c[8]};
const signed char **schar3DArr[2] = {schar3DArr0,schar3DArr1};
const unsigned int *uint1DArr = (const unsigned int*)&c[1];
const int *int1DArr = (const int*)&c[21];
const unsigned int *uint2DArr[2] = {(const unsigned int*)&c[1],(const unsigned int*)&c[17]};
const int *int2DArr[2] = {(const int*)&c[21],(const int*)&c[4]};
const unsigned int *uint3DArr0[2] = {(const unsigned int*)&c[1],(const unsigned int*)&c[17]};
const unsigned int *uint3DArr1[2] = {(const unsigned int*)&c[0],(const unsigned int*)&c[29]};
const unsigned int **uint3DArr[2] = {uint3DArr0,uint3DArr1};
const int *int3DArr0[2] = {(const int*)&c[21],(const int*)&c[4]};
const int *int3DArr1[2] = {(const int*)&c[10],(const int*)&c[25]};
const int **int3DArr[2] = {int3DArr0,int3DArr1};
```

Compacted and uncompacted data is used in exactly the same way, as shown in the following C-code:

```c
for (x=0; x<arrayLen; ++x){
  for (y=0; y<arrayLen; ++y){
    for (z=0; z<arrayLen; ++z){
      printf("%i ", int3DArr[x][y][z]);
    }
  }
}
```

The example for NULL pointers shows that they are added at the highest possible dimension allowing efficient handling of sparse matrices. Input data in pseudo-code notation:

```
uchar3DArrWithNull = [[NULL,[8, 16]],NULL]
```

Generated C-code with compacted data:

```c
#define NULL 0
const unsigned char *uchar3DArrWithNull0[2] = {NULL,(const unsigned char*)&c[37]};
const unsigned char **uchar3DArrWithNull[2] = {uchar3DArrWithNull0,NULL};
```

Here is an example C-code how to use these arrays:

```c
for (x=0; x<arrayLen; ++x){
  if (uchar3DArrWithNull[x]!=NULL){
    for (y=0; y<arrayLen; ++y){
      if (uchar3DArrWithNull[x][y]!=NULL){
        for (z=0; z<arrayLen; ++z){
          printf("%u ", uchar3DArrWithNull[x][y][z]);
        }
      }
    }
  }
}
```

### 3.3. Optimization

One way to improve the compaction result of dataCompactor is to choose different strategies for the greedy algorithm. Much more influence on the result is achieved by changing the size of the input array. Split Arrays (e.g. half size or less) can lead to a much better compaction rate for large arrays. Of course the modified array sizes have to be taken into account when accessing the data.

## 4. Comparison with Compression

To give a first impression of how data compaction performs against compression, four examples are given (see figure 1). For the comparison the python interfaces to the compression libraries "bz2" [5] and "zlib" [6] were used. The tool dataCompactor was applied with the methods "Remove Sub-Arrays" (listed as "1") and "Greedy Algorithm" (listed as "2"). Again two's complement for negative values is used with two bytes (A-C) and four bytes (D) big-endian representation for integer. The example includes the C-types `unsigned char`, `signed char`, `unsigned int`, and `int` with dimensions 1-3. Times were measured on a machine with an Intel Core i7-2620M CPU at 2.70GHz.

|  | bz2 compresslevel 9 | zlib compresslevel 9 | data compaction 1 | data compaction 1+2 |
|---|---:|---:|---:|---:|
| **Example A: number of input arrays: 28, Input size: 84 Bytes** | | | | |
| Compressed to percent | 109,52 | 72,62 | 57,14 | 48,81 |
| Compression time in ms | 0,03 | 0,01 | 0,11 | 2,79 |
| **Example B: number of input arrays: 228, Input size: 8084 Bytes** | | | | |
| Compressed to percent | 6,79 | 4,92 | 38,69 | 18,53 |
| Compression time in ms | 1,18 | 0,35 | 3,59 | 5609,59 |
| **Example C: number of input arrays: 3551, Input size: 87896 Bytes** | | | | |
| Compressed to percent | 13,08 | 10,09 | 20,69 | 18,01 |
| Compression time in ms | 32,36 | 4,11 | 298,94 | 2.803.969,29 |
| **Example D: number of input arrays: 2046, Input size: 151316 Bytes** | | | | |
| Compressed to percent | 4,53 | 4,32 | 26,28 | 24,62 |
| Compression time in ms | 12,62 | 12,12 | 160,66 | 646.234,45 |

**Figure 1:** Comparing compression size and time.

When looking at the compression times it must be considered that a proof of concept implementation is compared to highly optimized libraries. Nevertheless the greedy algorithm does not scale well, since it runs in super-linear.

Example C contains more arrays with overlap than example D. Therefore the greedy algorithm terminates faster for D with less additional memory reduction compared to removing sub-arrays. But overall in C and D the additional greedy algorithm does not reduce the compression size significantly. This of course depends on the input data. In D for example the input data consists of several parameter sets for software variants resulting in some identical arrays for different variants. So for some type of applications the results of the remove sub-arrays algorithm might be sufficient without the need to apply the time consuming greedy algorithm.

The small number of examples gives only a first impression about the compression rate of data compaction: it does not achieve the results of the compression libraries for larger data.

Please note that on a real embedded system the gap between the compression rates should be smaller: the additional memory consumption of "bz2" and "zlib" for the decompressed data (for data streams at least size of the longest array) as well as for the decompression algorithm itself is not taken into account in these examples, since they depend on a concrete implementation for an embedded system.

The conclusion for these first examples is that data compaction should be used if:
- There is a need for reducing the memory consumption.
- The optimization of computational or memory costs when accessing the data is more important than obtaining the best compression results.

# 5. Conclusions and Future Work

This paper introduces data compaction, a new approach for lossless and lossy compression of read-only array data. Data compaction overcomes the drawback of all existing compression approaches: computational and memory costs for decompression. This makes data compaction most suitable for systems that could currently not apply compression techniques due to real-time or memory constraints, i.e. for a wide range of embedded systems and for some multi-core applications (e.g. graphics cards).
No source code modification is needed to access compacted data. Moreover it can be integrated transparent into the compiling process. In addition data compaction supports efficient representations of arrays containing NULL pointers.
First results in comparison to existing compression libraries showed that data compaction performs not as well as existing approaches in terms of compression time and compression rate for larger data. Please note again that the gap between the compression rates of standard algorithms and data compaction should be smaller on real embedded systems (see section 4.). Data compaction is recommended if the optimization of computational or memory costs when using the data is more important than obtaining the best possible memory consumption.
For standard compression algorithms like "bz2" and "zlib" a reduction to about 1/4 percent of the original data size (see figure 1.) may not be considered as a good result. But most embedded systems can not apply these standard algorithms at all due to the reasons mentioned above. For these systems a memory reduction like this offers completely new possibilities:

- Use the gained memory for the implementation of additional features.
- Cost reduction by using less or smaller memory chips. Even small savings can sum up to significantly lower production costs, since embedded systems are typically sold in large quantities.

With dataCompactor a proof of concept implementation is presented. A more efficient implementation (e.g. by applying parallel programming) supporting more types (e.g. `float`) and dimensions (>3) would allow to apply data compaction for further application domains.
The approach was shown for standalone programs, further research could be done for data communication. Also code generator or compiler vendors could include data compaction in their tools. This would result in smaller programs without any effort from application developers.